\documentclass[10pt]{iopart}
\usepackage{amssymb}
\usepackage{amsfonts}
\usepackage{graphicx}

\begin{document}

\title{Renormalized spin coefficients in the accumulated orbital phase for
unequal mass black hole binaries}
\author{L\'{a}szl\'{o} \'{A}. Gergely$^{1,2\star }$, Peter L. Biermann$%
^{3,4,5,6,7\dag }$, \\
%EndAName
Bal\'{a}zs Mik\'{o}czi$^{8\P }$, Zolt\'{a}n Keresztes$^{1,2\ddag }$ }

\begin{abstract}
We analyze galactic black hole mergers and their emitted gravitational
waves. Such mergers have typically unequal masses with mass ratio of the
order $\,1/10$. The emitted gravitational waves carry the imprint of spins
and mass quadrupoles of the binary components. Among these contributions, we
consider here the quasi-precessional evolution of the spins. A method of
taking into account these third post-Newtonian (3PN) effects by
renormalizing (redefining) the 1.5 PN and 2PN accurate spin contributions to
the accumulated orbital phase is developed.
\end{abstract}

\bigskip

\address{
$^{1}$ Department of Theoretical Physics, University of Szeged, Tisza Lajos krt 84-86, Szeged 6720, Hungary\\
$^{2}$ Department of Experimental Physics, University of Szeged, D\'{o}m t\'{e}r 9, Szeged 6720, Hungary\\
$^{3}$ Max Planck Institute for Radioastronomy, Bonn, Germany\\
$^{4}$ Department of Physics and Astronomy, University of Bonn, Germany\\
$^{5}$ Department of Physics and Astronomy, University of Alabama,
Tuscaloosa, AL, USA \\
$^{6}$ Department of Physics, University of Alabama at Huntsville, AL, USA \\
$^{7}$ FZ Karlsruhe and Physics Department, University of Karlsruhe, Germany\\
$^{8}$ KFKI Research Institute for Particle and Nuclear Physics, Budapest 114, P.O.Box 49, H-1525 Hungary \\
$^{\star }$gergely@physx.u-szeged.hu 
$^{\dag }$plbiermann@mpifr-bonn.mpg.de 
$^{\P }$mikoczi@rmki.kfki.hu
$^{\ddag }$zkeresztes@titan.physx.u-szeged.hu}

\section{Introduction}

Mankind has admired the sky with its Milky Way, stars, planets, moons and
the Sun, the meteors and the Northern Light. In the age of satellites we map
the sky through all electromagnetic frequencies and also through high energy
cosmic particles. Gravitational waves are the last great frontier to be
reached and passed for the search for our physical understanding of the
universe via the messengers reaching us with the speed of light or very
close to it:

Gravitational waves, these fleeting distortions of space, the signature of
accelerating masses in the universe, remain to be discovered. Therefore it
is of paramount importance to understand, which objects are about to
generate strong gravitational waves.

When the black holes residing in the centers of almost all galaxies merge,
following the merger of their host galaxies, relatively strong gravitation
waves are emitted and various aspects of this process were already
investigated \cite{Peters}-\cite{spinflip}. A recent review of the generic
aspects of these galaxies nuclei as sources for ultra high energy cosmic
rays is in \cite{AGN}.

Starting from the broken powerlaw mass distribution of galactic central
black holes \cite{PressSchechter}-\cite{Ferrarese}, with simple assumptions
on the cross section and relative velocity of the two galaxies, we have
shown \cite{spinflip} that the most likely mass ratio $\nu =m_{2}/m_{1}\leq
1 $ of the merging black holes is between $1/3$ and $1/30$, thus a typical
value will be at about$\ 1/10$. Within this range of mass ratios the key
stages of the evolution leading to and following the merger can all be
treated with post-Newtonian (PN) approximations. A numerical analysis of
such mass ratios was recently presented in \cite{num10}.

In the next section we describe the sequence of events which will accompany
the merger of two supermassive galactic black holes. Then from Section 3 we
concentrate on the gravitational radiation dominated dissipative regime. An
important quantity in gravitational wave detection is the \textit{number of
cycles spent in the frequency range of the detector}. This quantity is
proportional to the \textit{accumulated orbital phase} (the total angle
covered by the quasi-circular orbital evolution). The gravitational waves
emerging from this process carry the imprint of the physical characteristics
(finite size) of the black holes, like their spins and quadrupole moments.
Spin precessions allow for increased accuracy and degeneracy breaking among
the parameters characterizing the source \cite{LH}.

We remark that according to \cite{RenormSigma}, the mass quadrupole - mass
monopole coupling for maximally rotating black holes (with quadrupole moment
originating entirely in their rotation) gives rise to contributions in the
accumulated orbital phase larger by a half to one order of magnitude than
the spin-spin contribution, at least for the typical unequal mass ratios.
Therefore for a successful detection this qualifies as the second most
important finite mass contribution to be considered, after the spin-orbit
effect.

In Section 3 we discuss the effect of the quasi-precessional evolution of
the spins on the accumulated phase of the gravitational waves and we present
estimates of the PN order in which various finite size contributions to the
change in the relative configurations of the spins and orbital angular
momentum contribute. The cumbersome equations leading to these estimates are
presented in the Appendix.

Sections 4 and 5 explore a convenient analytical way to encompass the effect
of spin and quadrupolar quasi-precessions even at a lower level accuracy.
The leading order contribution to the spin-spin (Section 4) and spin-orbit
(Section 5) contributions to the accumulated orbital phase are discussed
both for the equal mass and unequal mass cases in distinct subsections.

We summarize our findings in the Concluding Remarks.

\section{The merger of two supermassive galactic black holes: an odyssey}

First the dance: Two galaxies with central black holes approach each other
to within a distance where dynamical friction keeps them bound, spiraling
into each other. In the case when both galaxies with their central black
holes were radio galaxies, their jets get distorted and form the Z-shape 
\cite{Z}: the radio emitting blobs and tails produce the appearance of
dancing veils.

Second the meeting of the eyes: The central regions in each galaxy begin to
act as one unit, in a sea of stars and dark matter of the other galaxy.
During this phase, the central region can be stirred up, and produce a
nuclear starburst.

Third the lock: The black holes begin to lose orbital angular momentum due
to the interaction with the nearby stars \cite{Zier06}, and next by
gravitational radiation. The spin axes tumble and precess. This phase can be
identified with the apparent superdisk, as the rapidly precessing jet
produces the hydrodynamic equivalent of a powerful wind, by entraining the
ambient hot gas, pushing the two radio lobes apart and so giving rise the a
broad separation \cite{Gopal-Krishna07}. Due to the combined effect of
precessions and orbital angular momentum dissipation by gravitational waves
the spin (and hence jet) direction of the dominant black hole is reoriented
approximately in the direction of the original orbital angular momentum,
leading to a spin-flip \cite{spinflip}.

Fourth the plunge: The two black holes actually merge, with not much angular
orbital momentum left to be radiated away, thus the direction of the spin
remained basically unchanged (in the case of the typical mass ratio range we
discuss; for other setups the final spin may vary \cite{fS1}-\cite{BHcol}).
The final stage in this merger leads to a rapid increase in the frequency of
the gravitational waves, called \textquotedblleft chirping", but this
chirping will depend on the angles involved.

Fifth the rejuvenation: Now the newly oriented more massive merged black
hole starts its accretion disk and jet anew, boring a new hole for the jets
through its environment (the GigaHertz peaked sources). In this stage the
newly active jet is boring through the interstellar clouds with a gigantic
system of shockwaves, that accelerate protons and other charged particles in
a relativistic tennis-game. These very energetic particles then interact in
those same clouds, that slow down the progression of the relativistic jet,
and can so give rise to an abundance of neutrinos, TeV photons from pion
decay, and yet other particles and photons. These particles and their
interactions reach energies far beyond any Earth-bound accelerator.

Sixth the phoenix: The newly oriented jets begin to show up over some kpc,
and this corresponds to the X-shaped radio galaxies, while the old jets are
fading but still visible \cite{Rottman01}. This also explains many of the
compact steep spectrum sources, with disjoint directions for the inner and
outer jets.

Seventh the cocoon: The old jets have faded, and are at most visible in the
low radio frequency bubbly structures, such as seen for the Virgo cluster
region around M87. The feeding is slowing down, while a powerful jet is
still there, fed from the spin of the black hole.

Eighth the wait: The feeding of the black hole is down to catching some gas
out of a common red giant star wind as presumably is happening in our
Galactic center. This stage seems to exist for all black holes, even at very
low levels of activity. This black hole and its galaxy are ready for the
next merger, with the next black hole suitor, which may or may not come.

We predict that the super-disk radio galaxies should have large outer
distortions in their radio images, and should be visible in low keV X-rays.
A detection of the precessing jet, and its sudden weakening, would then
immediately precede the actual merger of the two black holes, and so may be
a predictor of the gravitational wave signal.

\section{The contribution of the spin precessions to the accumulated orbital
phase}

In the rest of the paper we concentrate on the gravitational radiation
dominated dissipative regime, when the supermassive black hole binary
radiates away gravitational waves.

If the binary system evolves unperturbed, the orbit circularizes faster than
it shrinks in radius. Therefore we consider circular orbits for which the
wave frequency is twice the orbital frequency.

Besides the Keplerian contribution to the accumulated orbital phase there
are general relativistic contributions, which scale with the post-Newtonian
(PN) parameter $\varepsilon \equiv m/r\approx v^{2}$ (with $m$ the total
mass and $r$ the radius of the reduced mass orbit about the total mass; we
take $G=1=c$). It is generally agreed from comparison with numerical
evolution, that all contributions up to 3.5 PN\ orders have to be taken into
account. To this order, beside the general relativistic contributions
starting at 1PN, there are various contributions originating in the finite
size of the binary components. These are related either to their rotation
(spin) or to their irregularities in shape (like a mass quadrupole moment).

Due to the continued accretion of surrounding matter into the supermassive
galactic black holes they spin up considerably. By taking into account only
the angular momentum transfer from accreting matter, the dimensionless spin
parameters $\chi _{i}=S_{i}/m_{i}^{2}$ (with $i=1,2$) grow to the maximally
allowed value $1$ even for initially non-rotating black holes \cite{Baardeen}%
. By taking into account the torque produced by the energy input of the
in-falling (horizon-crossing) photons emitted from the steady-state thin
accretion disk, which counteracts the torque due to mass accretion, the
limiting value of the dimensionless spin parameter is slightly reduced to $%
0.9982$ \cite{PageThorne}. Various refinements of this process, with the
inclusion of open or closed magnetic field lines \cite{MagneticAccretion}-%
\cite{WXL}, also jets in the magnetosphere of the hole \cite%
{JetDiskSymbiosys}-\cite{KovacsBiermannGergely} have not changed essentially
this prediction. Therefore we assume for the present considerations maximal
rotation $\chi _{i}\approx 1$. Due to this rotation the black hole is
centrifugally flattened (it becomes an oblate spheroid), a deformation which
can be characterized by a mass quadrupole scalar $Q_{i}$ \cite{Poisson}, or
its dimensionless counterpart $p_{i}=$ $Q_{i}/m_{i}m^{2}\approx -\left(
m_{i}/m\right) ^{2}$.

The accumulated orbital phase can be formally given as the PN expansion \cite%
{RenormSigma}, \cite{phi3PN3.5PN}, \cite{MVG}%
\begin{equation}
\phi =\phi _{c}+\phi _{N}+\phi _{1PN}+\phi _{1.5PN}+\phi _{2PN}+\phi
_{2.5PN}+\phi _{3PN}+\phi _{3.5PN}~.
\end{equation}%
Spin-orbit (SO) contributions appear at 1.5 PN, spin-spin (SS; composed of
proper S$_{1}$S$_{2}$ and self-contributions S$_{1}^{2}$ and S$_{2}^{2}$)
and mass quadrupole - mass monopole coupling (QM) contributions at 2PN. They
come with various PN corrections (PNSO at 2.5 PN; PNSS, PNQM and SO$^{2}$ at
3PN; finally 2PNSO at 3.5 PN; these were not computed yet with the exception
of PNSO \cite{PNSO2}-\cite{PNSO}). At 3PN there are additional spin and
quadrupole contributions, originating in the quasi-precessional evolution of
the spin and orbital angular momentum vectors. The number of gravitational
wave cycles $\mathcal{N}$ can be computed from the accumulated orbital phase
as $\mathcal{N}=\left( \phi _{c}-\phi \right) /\pi $ (the gravitational wave
frequency being twice the orbital frequency for quasi-circular motion).

The spin precession equations with SO, SS and QM contributions were given by
Barker and O'Connell \cite{BOC}. The relative geometry of the two spins $%
\mathbf{S}_{\mathbf{i}}$ and orbital angular momentum $\mathbf{L}$ can be
best described by the angles $\gamma =\arccos (\widehat{\mathbf{S}}_{\mathbf{%
1}}\cdot \widehat{\mathbf{S}}_{\mathbf{2}})$ and $\kappa _{i}=\arccos (%
\widehat{\mathbf{S}}_{\mathbf{i}}\cdot \mathbf{\hat{L}_{N})}$ (an overhat
denotes the direction of the respective vector). These angles are related by
the the spherical cosine identity%
\begin{equation}
\cos \gamma =\cos \kappa _{1}\cos \kappa _{2}+\cos \Delta \psi \sin \kappa
_{1}\sin \kappa _{2}\ ,  \label{gombi}
\end{equation}%
with $\Delta \psi $ the relative azimuthal angle of the spins. Due to
quasi-precessions, $\kappa _{i}$ and $\gamma $ evolve. Supplementing the SO
and SS contributions already given \cite{GPV} with the QM contributions
(with the quadrupole moment arising from pure rotation), we give these
expressions to second order accuracy:%
\begin{eqnarray}
(\cos \mathbf{\skew{39}{\dot}{\kappa}}_{i})\! &=&(\cos \mathbf{%
\skew{39}{\dot}{\kappa}}_{i})\!_{SO}+(\cos \mathbf{\skew{39}{\dot}{\kappa}}%
_{i})\!_{SS}+(\cos \mathbf{\skew{39}{\dot}{\kappa}}_{i})\!_{QM}\ , \\
(\cos \mathbf{\skew{27}{\dot}{\gamma}}\mathbf{)\!} &=&(\cos \mathbf{%
\skew{27}{\dot}{\gamma}}\mathbf{)\!}_{SO}+(\cos \mathbf{\skew{27}{\dot}{%
\gamma}}\mathbf{)\!}_{SS}+(\cos \mathbf{\skew{27}{\dot}{\gamma}}\mathbf{)\!}%
_{QM}~,
\end{eqnarray}%
with the detailed contributions enlisted in the Appendix.

The order of magnitude estimates of the terms in Eqs. (\ref{kappa1_SO})-(\ref%
{gamma_QM}) (cf. a footnote in Ref. \cite{spinspin2}, according to which $%
O(\delta x)=O(\dot{x})r\varepsilon ^{-1/2}$), and assuming maximal rotation
lead to%
\begin{eqnarray}
\mathcal{O}\left( \delta _{SO}\kappa _{1}\right) &\approx &\varepsilon
^{3/2}\left( 1+\nu ^{-1}\right) ^{-2}{\left( 2+\nu ^{-1}\right) }\approx
\varepsilon ^{3/2}\nu \ ,  \nonumber \\
\mathcal{O}\left( \delta _{SS}\kappa _{1}\right) &\approx &\varepsilon
^{3/2}\left( 1+\nu ^{-1}\right) ^{-2}(1+\varepsilon ^{1/2}\nu ^{-1})\approx
\varepsilon ^{3/2}\nu (\nu +\varepsilon ^{1/2})\ ,  \nonumber \\
\mathcal{O}\left( \delta _{QM}\kappa _{1}\right) &\approx &\varepsilon
^{3/2}(\nu \mathbf{+}\varepsilon ^{1/2}\nu ^{2})\approx \varepsilon
^{3/2}\nu \ ,  \nonumber
\end{eqnarray}%
\begin{eqnarray}
\mathcal{O}\left( \delta _{SO}\kappa _{2}\right) &\approx &\varepsilon
^{3/2}\left( 1+\nu \right) ^{-2}{\left( 2+\nu \right) \approx }\varepsilon
^{3/2}\ ,  \nonumber \\
\mathcal{O}\left( \delta _{SS}\kappa _{2}\right) &\approx &\varepsilon
^{3/2}\left( 1+\nu \right) ^{-2}(1-\varepsilon ^{1/2}\nu )\approx
\varepsilon ^{3/2}\ ,  \nonumber \\
\mathcal{O}\left( \delta _{QM}\kappa _{2}\right) &\approx &\varepsilon
^{3/2}(\nu \mathbf{+}\varepsilon ^{1/2})\approx \varepsilon ^{3/2}(\nu 
\mathbf{+}\varepsilon ^{1/2})\ ,  \nonumber
\end{eqnarray}%
\begin{eqnarray}
\mathcal{O}\left( \delta _{SO}\gamma \right) &\approx &\varepsilon \left(
2+\nu +\nu ^{-1}\right) ^{-1}\left( \nu -\nu ^{-1}\right) \approx
\varepsilon \ ,  \nonumber \\
\mathcal{O}\left( \delta _{SS}\gamma \right) &\approx &\varepsilon ^{3/2}%
\left[ \left( 1+\nu ^{-1}\right) ^{-2}-\left( 1+\nu \right) ^{-2}\right]
\approx \varepsilon ^{3/2}\ ,  \nonumber \\
\mathcal{O}\left( \delta _{QM}\gamma \right) &\approx &\varepsilon
^{3/2}\left( -\nu +\nu \right) \approx \varepsilon ^{3/2}\nu \ .
\label{orders}
\end{eqnarray}%
Thus the angle $\gamma $ varies at 1PN, while the angles $\kappa _{i}$ only
at 1.5 PN \textit{during one orbital period}. However $\gamma $ appears only
in the 2PN order S$_{1}$S$_{2}$ contribution to the accumulated orbital
phase as opposed to the angles $\kappa _{i}$ which are present in all
finite-size contributions, in particular in the 1.5 PN order SO contribution.

Therefore 3PN order quasi-precessional contributions to the accumulated
orbital phase in the comparable mass case $\nu \lesssim 1$ arise from $%
\delta _{SO}\kappa _{i}$, $\delta _{SS}\kappa _{i}$, $\delta _{QM}\kappa
_{i} $ and $\delta _{SO}\gamma $; while in the unequal mass case, found
typical for supermassive galactic black hole binaries only from $\delta
_{SO}\kappa _{2}$, $\delta _{SS}\kappa _{2}$ and $\delta _{SO}\gamma $.

\section{Leading order evolution of the spin-spin coefficient}

From among the quasi-precessional evolutions, we have recently integrated 
\cite{RenormSigma} the SO evolution (averaged over one orbit) of the angle $%
\Delta \psi $, finding%
\begin{equation}
\Delta \psi =\left( \Delta \psi \right) _{0}+\frac{3\mu n\left( \nu
^{-1}-\nu \right) }{2a}t~.  \label{DePsi}
\end{equation}%
Here~$\mu =m_{1}m_{2}/m$ the reduced mass, $a$ the radius and $n=2\pi
/T_{orbit}=\pi /T_{wave}$. The time-dependent expression for $\gamma $ is
then given by Eq. (\ref{gombi}), with $\Delta \psi $ from Eq. (\ref{DePsi}).
The variation of both $\Delta \psi $ and $\gamma $ thus have a periodicity
with period%
\begin{equation}
T_{3PNS}=F\left( \nu \right) \varepsilon ^{-1}T_{wave}~,  \label{T3PNS}
\end{equation}%
where 
\begin{equation}
F\left( \nu \right) =\frac{4\left( 2+\nu +\nu ^{-1}\right) }{3\left( \nu
^{-1}-\nu \right) }~.
\end{equation}

\subsection{Equal masses}

For equal mass binaries this time-scale becomes infinite $\lim_{\nu
\rightarrow 1}F\left( \nu \right) =\infty $ (see Fig1a) expressing the fact
that the time-dependent part of $\gamma $ goes to zero. Thus $\gamma $ is a
constant and \textit{there is no 3PN quasi-precessional contribution to the
accumulated orbital phase for equal mass binaries}.

\subsection{Unequal masses}

For the typical mass ratio the factor $F\left( \nu \right) $~is of the order
of unity (see Fig 1b), thus the variation time-scale of $\gamma $ is $%
\varepsilon ^{-1}\in \left( 10,1000\right) $ times larger than the wave
period. The contribution of the quasi-precessions can be taken into account
then by keeping only the first term in Eq. (\ref{gombi}) when computing the
2PN order S$_{1}$S$_{2}$ contribution to the accumulated orbital phase. e.
g. by introducing a \textit{renormalized spin coefficient}. This coefficient
arises by modifying the 2PN contribution with the average of some of the 3PN
contributions. By using this renormalized coefficient in the context of the
2PN accurate dynamics, one can bring closer the 2PN analytical prediction to
the numerical results \cite{RenormSigma}.\footnote{%
However we stress that there are other finite size 3PN order contributions
to the phase, like PNSS, PNQM and SO$^{2}$, which are still not computed.}

Thus, in the unequal mass case we propose to replace 
\begin{equation}
\sigma _{S_{1}S_{2}}=\frac{S_{1}S_{2}}{48\eta m^{4}}(-247\cos \gamma
+721\cos \kappa _{1}\cos \kappa _{2})~,
\end{equation}%
with 
\begin{equation}
\overline{\sigma _{S_{1}S_{2}}}=\frac{79S_{1}S_{2}}{8\eta m^{4}}\cos \kappa
_{1}\cos \kappa _{2}\ ,
\end{equation}%
in the respective 2PN\ contribution to the accumulated orbital phase 
\begin{eqnarray}
\phi _{2PN} &=&-\frac{1}{\eta }\left( \frac{9275495}{14450688}+\frac{284875}{%
258048}\eta +\frac{1855}{2048}\eta ^{2}-\frac{15}{64}\sigma \right) \tau
^{1/8}~,  \nonumber \\
\sigma &=&\sigma _{S_{1}S_{2}}+\sigma _{SS-self}+\sigma _{QM}\ .
\end{eqnarray}%
(Here $\eta =\mu /m$ is the symmetric mass ratio and $\tau $ a dimensionless
time parameter.)

\begin{figure}[t]
\includegraphics[height=3.7cm]{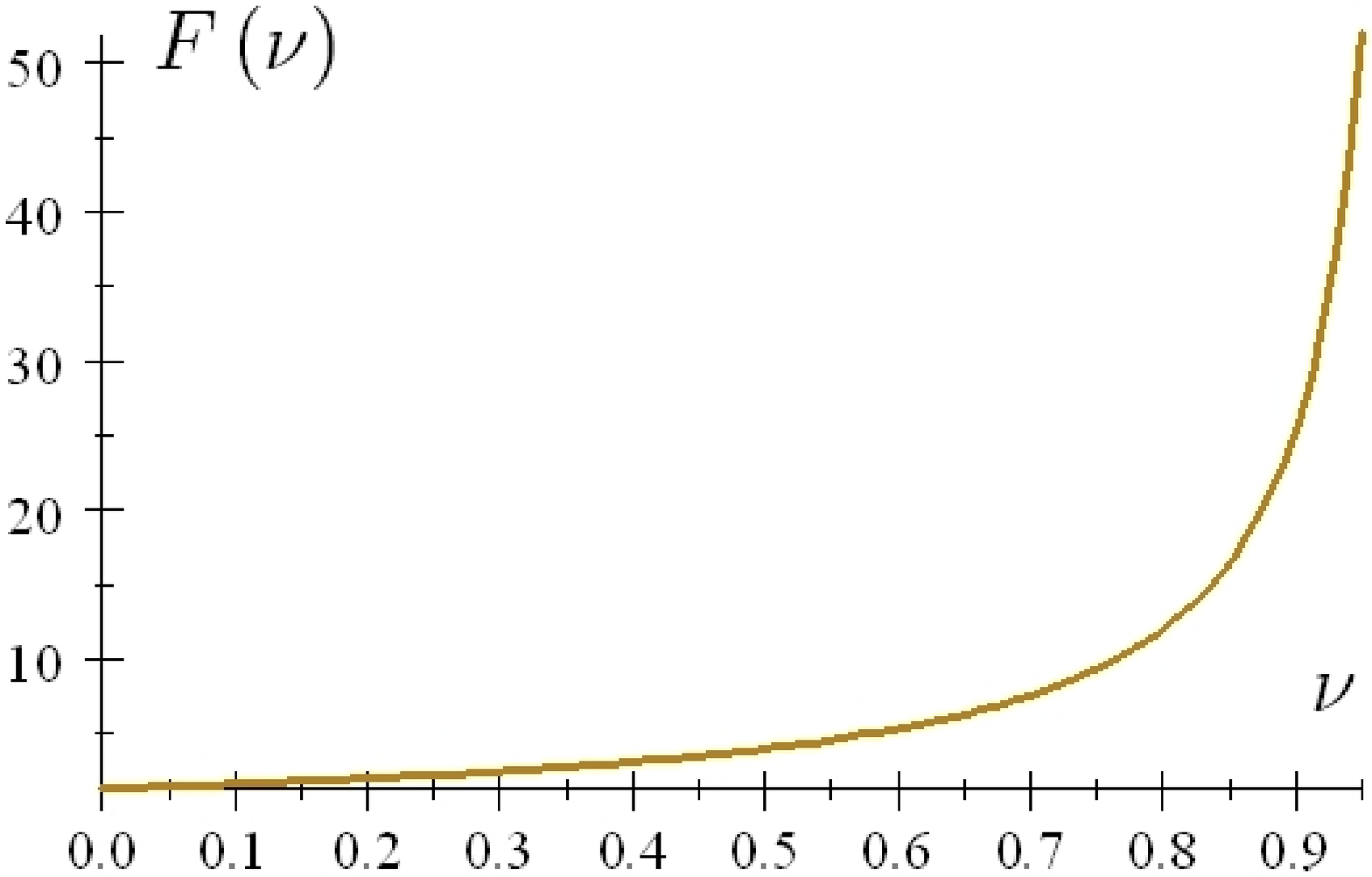} \hskip1.3cm %
\includegraphics[height=3.7cm]{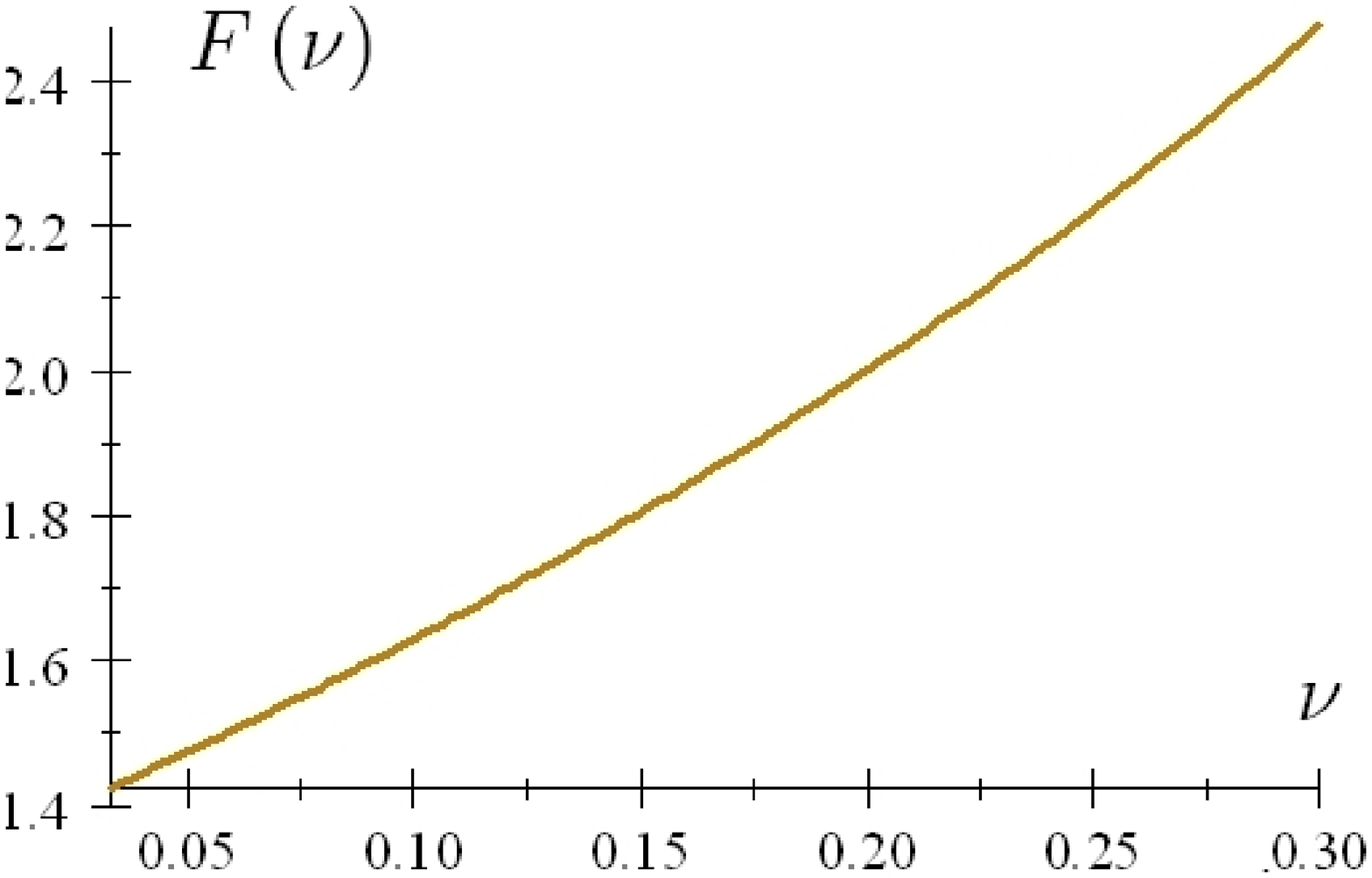}
\caption{(Color online) (a) The factor $F\left( y\right) $, shown for the
whole allowed range of mass ratios, diverges in the equal mass case. (b) The
factor $F\left( y\right) $ in the typical mass ratio range $\left(
1/30,~1/3\right) $ found for merging supermassive galactic black hole
binaries is of order unity.}
\label{Fig1}
\end{figure}

\section{Leading order evolution of the spin-orbit coefficient}

The 1.5 PN contribution to the accumulated orbital phase is 
\begin{equation}
\phi _{1.5PN}=-\frac{3}{4\eta }\left( \frac{1}{4}\beta -\pi \right) \tau
^{1/4}~,
\end{equation}%
with the SO contribution given by 
\begin{equation}
\beta =\frac{1}{12}\sum_{i=1}^{2}\chi _{i}\cos \kappa _{i}\left( 113\frac{%
m_{i}^{2}}{m^{2}}+75\eta \right) \ .  \label{SO}
\end{equation}

\subsection{Equal masses}

The quasi-precessional equations with the SO, SS and QM contributions
included were recently integrated by Racine for the equal mass case \cite%
{Racine}. According to his analysis, there is a previously unknown constant
of the motion%
\begin{equation}
\lambda =\frac{2}{L_{N}}\left( S_{1}\cos \kappa _{1}+S_{2}\cos \kappa
_{2}\right) ~.
\end{equation}%
As for equal masses $\beta =\left( 47L_{N}/24\right) \lambda =$const., 
\textit{the quasi-precessional evolution at 3PN does not affect the
coefficient }$\beta $\textit{\ up to the 3PN accuracy.}

\subsection{Unequal masses}

In the unequal mass case the leading order evolutions of the angles $\kappa
_{i}$ are $\delta _{SO}\kappa _{2}$ and $\delta _{SS}\kappa _{2}$. Averaged
over one orbit they give \cite{GPV}%
\begin{equation}
\dot{\kappa}_{2}=\frac{3\left( 1+\nu \right) m_{1}^{2}}{2a^{3}}\sin \kappa
_{1}\sin \Delta \psi ~.
\end{equation}%
For unequal masses $\nu \approx 10^{-1}$ the estimates (\ref{orders}) allow
to consider $\kappa _{1}$ as a constant (in a first approximation). The
leading-order time-dependence of $\Delta \psi $ is given by Eq. (\ref{DePsi}%
). Integration gives 
\begin{equation}
\kappa _{2}=\left( \kappa _{2}\right) _{0}-\frac{\varepsilon ^{1/2}}{\left(
1-\nu \right) }\sin \kappa _{1}\cos \left[ \left( \Delta \psi \right) _{0}+%
\frac{3\mu n\left( \nu ^{-1}-\nu \right) }{2a}t\right] ~.  \label{kappa2}
\end{equation}%
(We have employed $\varepsilon \approx m/a$ .) The time-varying part has the
same period $T_{3PNS}$ given by Eq. (\ref{T3PNS}), which is of the scale of $%
\varepsilon ^{-1}$ the orbital period. In agreement with this and the
estimate (\ref{orders}) 
\begin{equation}
\frac{\mathcal{O}\left( \kappa _{2}-\left( \kappa _{2}\right) _{0}\right) }{%
\mathcal{O}\left( \delta _{SO,SS}\kappa _{2}\right) }=\frac{\varepsilon
^{1/2}}{\varepsilon ^{3/2}}=\varepsilon ^{-1}~.
\end{equation}%
As the time varying part averages out, to leading order accuracy we can
simply identify the angle $\kappa _{2}$ with its constant initial value $%
\left( \kappa _{2}\right) _{0}$.

In conclusion, in the unequal mass case and in the leading order
approximation $\kappa _{1}=$const. the time variation due to
quasi-precessions renders $\kappa _{2}$ to another constant at 3PN accuracy. 
\textit{There is no need to renormalize }$\beta $, \textit{when taking into
account the leading order quasi-precessional contributions.}

\section{Concluding Remarks}

The process of the merger of two supermassive galactic black holes gives
rise to exciting signals in all the messengers accessible now and in the
near future, and will allow to probe some of the most profound secrets of
Nature, occurring at energies and circumstances far beyond anything on
Earth. We have described the sequence of events during the process of
merging of such supermassive galactic black holes with reference to related
processes involving accretion, jets, energetic particles and electromagnetic
signatures. We argued that the class of radio galaxies with a super-disk are
particularly promising candidates for future powerful low frequency
gravitational wave emission.

The accumulated orbital phase is a quantity to be known precisely for
successful detection of gravitational waves. Besides Keplerian, first and
second PN\ order general relativistic contributions, the finite size of the
black holes also contributes, first at 1.5 PN orders via the spin-orbit
interaction, then at 2PN by the spin-spin and quadrupole-monopole coupling.
These contributions are further modulated by quasi-precessional evolutions
of the spins. However this is but a subset of all possible finite size
contributions at 3PN, as stressed earlier. The way we propose to take into
account the precessional 3PN contributions is to add their average to the
lower order terms, a procedure we call \textit{renormalization }of the lower
order coefficients.

We have proven here that there are no such quasi-precessional contributions
in the equal mass case. Furthermore we have proved for the unequal mass case
that renormalizing the 1.5 PN spin-orbit coefficient will not change its
value. At 2PN there are quadrupolar and spin-spin contributions to consider.
As the former can be regarded as constant up to 3PN, the only modification
due to quasi-precessions comes from the spin-spin interaction. The
renormalized spin-spin coefficient at 2PN thus is adjusted by the average\
of the 3PN quasi-precessional contribution, taken over the period of change
induced by these quasi-precessions (a time-scale one PN order lower than the
orbital period). We have checked that the renormalized spin-spin coefficient
gives closer results to the numerical estimates for compact binaries with
the mass ratio of order $1/10$ \cite{RenormSigma}.

Our results related to the renormalization of the spin coefficients are
significant for supermassive black hole binaries, which are typically LISA
sources. However as we have employed nothing but the mass ratio of the
binary in our considerations, they also apply for binaries composed of an
intermediate mass and a stellar mass black hole, which qualify as LIGO
sources.

\ack Participation of L\'{A}G at the GWDAW13 meeting was supported by the Pol%
\'{a}nyi Program of the Hungarian National Office for Research and
Technology (NKTH). L\'{A}G was also successively supported by a ROF Award of
LSBU and by Collegium Budapest. Support for work with PLB has come from the
AUGER membership and theory grant 05 CU 5PD 1/2 via DESY/BMBF and VIHKOS.
The work of BM was supported by the Hungarian Scientific Research Fund
(OTKA) grant no. 68228; and of ZK by the OTKA grant no. 69036.

\appendix

\section{Quasi-precessional evolutions of the relative spin angles}

The SO and SS contributions of the instantaneous variations of the relative
spin angles \cite{GPV} supplemented with the respective QM contributions
(assuming the quadrupole moment arises from pure rotation), to second order
accuracy are:%
\begin{eqnarray}
(\cos \mathbf{\skew{44}{\dot}{\kappa}}_{1})\!_{\,SO} &=&\frac{3S_{2}}{2r^{3}}%
\left( 2+\nu ^{-1}\right) \mathbf{\hat{L}_{N}\cdot (\hat{S}_{1}\times \hat{S}%
_{2}})\ ,  \label{kappa1_SO} \\
(\cos \mathbf{\skew{44}{\dot}{\kappa}}_{1})\!_{\,SS} &=&\frac{3S_{2}}{r^{3}}%
\left[ (\mathbf{\hat{r}\cdot \hat{S}_{2}})\mathbf{\hat{L}_{N}\cdot }\left( 
\mathbf{\hat{r}}\times \mathbf{\hat{S}_{1}}\right) \right.  \nonumber \\
&&+\left. \frac{S_{1}}{L_{N}}(\mathbf{\hat{r}\cdot \hat{S}}_{\mathbf{1}})%
\mathbf{\hat{r}\cdot (\hat{S}_{1}\times \hat{S}_{2}})\right] \ ,
\label{kappa1_SS} \\
(\cos \mathbf{\skew{44}{\dot}{\kappa}}_{1})\!_{\,QM} &=&\mathbf{-}\frac{3\mu
m^{3}}{r^{3}}\left[ \frac{p_{1}}{S_{1}}(\mathbf{\hat{r}\cdot \hat{S}}_{%
\mathbf{1}})\mathbf{\hat{L}_{N}\cdot }(\mathbf{\hat{r}}\times \mathbf{\hat{S}%
_{1})}\right.  \nonumber \\
&&\mathbf{+}\left. \frac{p_{2}}{L_{N}}(\mathbf{\hat{r}\cdot \hat{S}}_{%
\mathbf{2}})\mathbf{\hat{r}\cdot (\hat{S}_{1}\times \hat{S}_{2}})\right] \ ,
\label{kappa1_QM}
\end{eqnarray}%
\begin{eqnarray}
(\cos \mathbf{\skew{44}{\dot}{\kappa}}_{2}\mathbf{)}_{\!SO} &=&-\frac{3S_{1}%
}{2r^{3}}{\left( 2+\nu \right) }\mathbf{\hat{L}_{N}\cdot }(\mathbf{\hat{S}%
_{1}}\times \mathbf{\hat{S}_{2})}\ ,  \label{kappa2_SO} \\
(\cos \mathbf{\skew{44}{\dot}{\kappa}}_{2}\mathbf{)}_{\!SS} &=&\frac{3S_{1}}{%
r^{3}}\left[ (\mathbf{\hat{r}\cdot \hat{S}_{1}})\mathbf{\hat{L}_{N}\cdot }(%
\mathbf{\hat{r}}\times \mathbf{\hat{S}_{2})}\right.  \nonumber \\
&&-\left. \frac{S_{2}}{L_{N}}(\mathbf{\hat{r}\cdot \hat{S}}_{\mathbf{2}})%
\mathbf{\hat{r}\cdot (\hat{S}_{1}\times \hat{S}_{2}})\right] \ ,
\label{kappa2_SS} \\
(\cos \mathbf{\skew{44}{\dot}{\kappa}}_{2}\mathbf{)}_{\!QM} &=&-\frac{3\mu
m^{3}}{r^{3}}\left[ \frac{p_{2}}{S_{2}}(\mathbf{\hat{r}\cdot \hat{S}_{2}})%
\mathbf{\hat{L}_{N}\cdot }(\mathbf{\hat{r}}\times \mathbf{\hat{S}_{2})}%
\right.  \nonumber \\
&&\mathbf{+}\left. \frac{p_{1}}{L_{N}}\mathbf{(\hat{r}\cdot \hat{S}_{1})\hat{%
r}\cdot (\hat{S}_{1}\times \hat{S}_{2}})\right] \ ,  \label{kappa2_QM}
\end{eqnarray}%
\begin{eqnarray}
(\cos \mathbf{\skew{27}{\dot}{\gamma}}\mathbf{)\!}_{SO} &=&\frac{3L_{N}}{%
2r^{3}}\left( \nu -\nu ^{-1}\right) \mathbf{\hat{L}_{N}\cdot }(\mathbf{\hat{S%
}_{1}}\times \mathbf{\hat{S}_{2})}\ ,  \label{gamma_SO} \\
(\cos \mathbf{\skew{27}{\dot}{\gamma}}\mathbf{)\!}_{SS} &=&\frac{3}{r^{3}}%
\left[ S_{2}(\mathbf{\hat{r}\cdot \hat{S}_{2}})-S_{1}(\mathbf{\hat{r}\cdot 
\hat{S}_{1}})\right] \mathbf{\hat{r}\cdot }(\mathbf{\hat{S}_{1}}\times 
\mathbf{\hat{S}_{2})}\ ,  \label{gamma_SS} \\
(\cos \mathbf{\skew{27}{\dot}{\gamma}}\mathbf{)\!}_{QM} &=&\frac{3\mu m^{3}}{%
r^{3}}\left[ \frac{p_{2}}{S_{2}}(\mathbf{\hat{r}\cdot \hat{S}_{2}})-\frac{%
p_{1}}{S_{1}}(\mathbf{\hat{r}\cdot \hat{S}_{1}})\right] \mathbf{\hat{r}\cdot 
}(\mathbf{\hat{S}_{1}}\times \mathbf{\hat{S}_{2})}\ .  \label{gamma_QM}
\end{eqnarray}

\end{document}